\documentclass[a4paper,twoside]{article}

\usepackage[utf8]{inputenc}%(only for the pdftex engine)
\usepackage{url}

\usepackage{epsfig}
\usepackage{subcaption}
\usepackage{calc}
\usepackage{amssymb}
\usepackage{amstext}
\usepackage{amsmath}
\usepackage{amsthm}
\usepackage{multicol}
\usepackage{multirow}
\usepackage{pslatex}
\usepackage{apalike}
\usepackage[bottom]{footmisc}
\usepackage{SCITEPRESS}     % Please add other packages that you may need BEFORE the SCITEPRESS.sty package.

\usepackage{siunitx}

\begin{document}

  \title{\huge Data Protection and Security Issues With Network Error Logging}

\author{\authorname{Libor Polčák\sup{1}\orcidAuthor{0000-0001-9177-3073} and Kamil Jeřábek\sup{1}\orcidAuthor{0000-0002-5317-9222}}
\affiliation{\sup{1}Brno University of Technnology, Faculty of Information
Technology, Božetěchova 2, 612 66 Brno, Czech Republic}
\email{\{polcak, ijerabek\}@fit.vut.cz}
}


  \abstract{
    Network Error Logging
    helps web server operators detect operational problems in real-time
    to provide fast and reliable services. This paper analyses Network Error
    Logging from two angles.
    Firstly, this paper overviews Network Error Logging from the data protection view.
    The ePrivacy Directive requires consent for non-essential access to the end
    devices. Nevertheless, the Network Error Logging design does not allow
    limiting the tracking to consenting users. Other issues lay in GDPR
    requirements for transparency and the obligations in the contract between
    controllers and processors of personal data.
    Secondly, this paper explains Network Error Logging exploitations to
    deploy
    long-time trackers to the victim devices. Even though
    users should be able to disable Network Error Logging, it is not clear how to
    do so. Web server operators can mitigate the attack by configuring servers to
    preventively remove policies that adversaries might have added.
}

\onecolumn \maketitle \normalsize \setcounter{footnote}{0} \vfill

\section{\uppercase{Introduction}}
\noindent Web server operators need to monitor their servers for
availability, which is a crucial success factor~\cite{franke2012}.
Network Error Logging (NEL)~\cite{w3c_nel} detects failures in the
reachability of web servers, including failures during domain name
resolution~\cite{paper_nel}. Each HTTP server can employ each visiting
browser\footnote{At the time of the writing of this paper, Chromium-based
browsers like Google Chrome, Chrome for Android, Microsoft Edge, Opera, and Opera
GF support NEL. Brave is the only Chromium-based browser that we discovered does not support NEL.}
for NEL by adding a NEL header into the HTTP reply. Consequently, the web server
operator can learn about future failed attempts of the browser to reach the web
server. NEL also allows to report of successful visits and provides a fallback
mechanism for unreachable logging servers.

The current NEL editor's standard draft~\cite{w3c_nel} covers various privacy implications of NEL.
For example, an
attacker can distribute
unique reporting URLs for each user; consequently, NEL can be misused as a
supercookie. The paper describing NEL~\cite{paper_nel} provides four security,
privacy, and ethical principles:
(1)~NEL logs existing information in a different place, (2)~NEL logs only requests
users voluntarily make, (3)~end users can opt out of NEL, and (4)~third parties must
not be able to use NEL to monitor sites they do not control.

This paper has two goals. Firstly,
we go through the law requirements of NEL in the European
Economic Area (EEA) and show that web server operators should seek consent
from users
before installing NEL policies to browsers. Secondly, we review
the security, privacy, and ethical expectations of
the original NEL paper~\cite{paper_nel} and check their correctness from the
user's perspective.

This paper has several contributions:

\begin{enumerate}

  \item We are the first to analyze NEL law requirements in EEA.

  \item We provide arguments disputing the security, privacy, and ethical
    expectation of NEL authors~\cite{paper_nel} not covered by the current NEL
    editor's standard draft~\cite{w3c_nel}.

  \item We are first to suggest that web server operators that do not deploy NEL
    should configure their servers to remove policies that adversaries might
    have added. \cite{nel_http_archive} revealed only a few domains
    removing NEL, so such configuration is not deployed in the wild.

\end{enumerate}

This paper is organized as follows. Section~\ref{sec:nelbackground} explains how
NEL works. Section~\ref{sec:related} overviews related
work.
Section~\ref{sec:law} analyzes the requirements stemming from the EEA law.
Section~\ref{sec:disputes} disputes the security, privacy, and ethical
  expectation of NEL.
Section~\ref{sec:discussion} discusses the implications of the results of this
  paper.
Section~\ref{sec:conclusion} concludes this paper.

\section{\uppercase{NEL Background}}
\label{sec:nelbackground}
\noindent NEL was introduced~ by researchers mostly affiliated with
Google~\cite{paper_nel}. The World Wide Web
Consortium (W3C) is in the process of standardizing
NEL~\cite{w3c_nel}.
NEL introduces {\tt NEL} HTTP header sent by an HTTP server that contains the NEL
policy of the server. In addition,
{\tt Report-To} HTTP header determines web servers that collect
NEL reports for the visited domain.

The goal of NEL is to let web server operators instruct their
visitors to report failures in attempting to visit the service as well as
successful visits.
A web server operator can control the fraction of failures
({\tt failure\_fraction})
and success reports ({\tt success\_fraction}), and other parameters in a NEL policy. The policy is valid
for a limited time ({\tt max\_age} parameter of a policy). By default, the policy
  applies  to the visited domains, but a server can also instruct the clients
  to apply the policy to subdomains ({\tt include\_subdomains} parameter of a policy).

A NEL collector can deploy NEL on its own. Hence, once a browser learns the NEL
policy of a domain and sends a report, it learns the NEL policy of the collector
(if deployed). Reports about the collector are called meta
reports~\cite{paper_nel}.

Figure~\ref{fig:nel_report_w3c} shows an example of a NEL report. The client
reports the age of the error (the browser often sends messages with a delay)
and its type. Additionally,
the report contains other information about the event so that an
operator can react to the message.

\begin{figure}[h]
    \centering
\begin{verbatim}
{
  "age": 0,
  "type": "network-error",
  "url": "https://www.example.com/",
  "body": {
    "sampling_fraction": 0.5,
    "referrer": "http://example.com/",
    "server_ip": "2001:DB8:0:0:0:0:0:42",
    "protocol": "h2",
    "method": "GET",
    "request_headers": {},
    "response_headers": {},
    "status_code": 200,
    "elapsed_time": 823,
    "phase": "application",
    "type": "http.protocol.error"
  }
}
\end{verbatim}
    \caption{An example of a NEL report~\cite{w3c_nel}.}
    \label{fig:nel_report_w3c}
\end{figure}

Browsers store only the
last policy retrieved for each domain.

\section{\uppercase{Related Work}}
\label{sec:related}
\noindent \cite{nel_http_archive} analyze HTTP Archive in a longitudinal study.
11.73\,\% (almost 2,250,000 unique domains) deployed NEL in February 2023.
Cloudflare is the largest NEL provider according to the study.
To our best knowledge, all other papers concerning NEL
utilize NEL as a tool. This paper is the first that
puts the NEL design decisions
under scrutiny. Nevertheless, the privacy considerations of the current NEL
editor's standard draft~\cite{w3c_nel} cover several issues that this paper does not repeat.

Online tracking is
omnipresent~\cite{Krishnamurthy2011,WebTrackingPolicyTechnology,session-replay,ico_adtech}.
Cookies holding unique identifiers may be accompanied or replaced with ETags and
other tracking mechanisms like browser fingerprinting~\cite{WebTrackingPolicyTechnology}.
Indeed, NEL can be misused for online tracking as covered by the privacy
considerations of the current NEL
editor's standard draft~\cite{w3c_nel}.

The original request-response nature of HTTP has already been
accompanied by other techniques that allow browsers to communicate with servers
without any explicit user action initiating such communication. For example,
Push API~\cite{w3c_push} allows a previously
visited server to push a message without the user needing to open the web page in the browser.
Nevertheless, the user must consent to allow a site to access Push API. Hyperlink
auditing\footnote{\url{https://html.spec.whatwg.org/multipage/links.html#hyperlink-auditing}} allows
web page creators to force the browser to report that the user clicked a link.

Previous research shows that 48--75\,\% of websites unintentionally leak
identifiers and other private information in URLs \cite{Krishnamurthy2011,WebTrackingPolicyTechnology}.
A NEL report contains both the URL of the previously visited page ({\tt referrer}) and the page
just visited (success) or broken page (error report), {\tt url}.
As {\tt Referer} HTTP header leaks the previously visited web page URL,
NEL would log the
same information as detected by~\cite{Krishnamurthy2011}.
Nevertheless, the {\tt Referer} header already went through scrutiny, and web pages can
regulate the content of the
header\footnote{\url{https://developer.mozilla.org/en-US/docs/Web/Security/Referer_header:_privacy_and_security_concerns}}.
NEL applies the same restrictions to the referrer field as the
{\tt Referer} header~\cite{w3c_nel}. % https://www.w3.org/TR/referrer-policy/
%We experimented with the {\tt Referer} header and NEL {\tt
%referrer} and the browser behaved consistently even when we installed an
%extension removing the information.

An adversary with access to the communication channel between a web visitor
and a web server accessed by the visitor can deploy a Man In The Middle (MitM)
attack~\cite{mitmsurvey}. As NEL can be deployed only on TLS
channels~\cite{w3c_nel,paper_nel}, an adversary would need to spoof a valid
certificate~\cite{mitmsurvey}, hope that the victim ignores the
error~\cite{ssl_effectiveness,kazakhstan_ca}, exploit a broken
implementation~\cite{https_mitm_vulnerabilities}, or remove the
encryption~\cite{hsts_survey}. Section~\ref{sec:discussion} builds on the idea
that MitM attacks appear in practice.

\section{\uppercase{Law Analysis}}
\label{sec:law}
\noindent In EEA, ePrivacy Directive regulates
publicly available services and networks~\cite{ePrivacy,ePrivacyAmmendment,vanhoboken2015}.
\emph{"Article 5(3)
applies to anyone that stores or accesses information, such as a cookie, on a
user's device, including if no personal data are
involved."}~\cite{vanhoboken2015}  Article 29 Working Party (WP29) also considers ePrivacy Directive to
apply to different technologies, not only
cookies~\cite[Introduction]{wp29_cookies_consent}. In essence, Article 5(3)
allows only strictly necessary access to the user's terminal device or the access is
needed to carry the communication~\cite{wp29_cookies_consent}. Specifically, the
necessity needs to be viewed from the point of the \emph{user}, not the service
provider~\cite[Section~5]{wp29_cookies_consent}.

Additionally, when personal data are processed, GDPR~\cite{gdpr} regulates the
processing unless ePrivacy Directive overrides the specific
provisions~\cite[Recital 74]{cjeuC645_19}.

NEL requires storing policies in the browser~\cite[Process policy headers, step
14]{w3c_nel}. Hence, ePrivacy Directive applies. Although web server operators
view NEL as beneficial~\cite{paper_nel}, it is not strictly necessary from the
point of the user.
One of the four principles of NEL
is that the user can opt out of NEL at any time~\cite[Section 3.2, point~3]{paper_nel}.
\cite{nel_http_archive} show that most web
servers do not need NEL to function.
Consequently, neither of the two exceptions of the ePrivacy
Directive applies. Hence, obtaining consent is necessary before installing a NEL
policy. As the current NEL implementations
do not obtain consent~\cite{paper_nel,w3c_nel}, the site operator have
to find another way to seek consent before adding NEL headers to its replies.

%Indeed, looking at the WP29 argumentation on
%analytics~\cite[Section 4.3]{wp29_cookies_consent}, similar arguments can be
%made on NEL.

Additionally, the information reported by NEL can consist of personal data:

\begin{enumerate}

  \item Personal data can be embedded in
    URLs~\cite{Krishnamurthy2011,WebTrackingPolicyTechnology,cnil_ga_onestopshop,UOOUReservatics}. Although
    the original NEL paper expects that NEL reports do not contain URL parameters~\cite[Section
    3.4]{paper_nel}, such a requirement is missing in the current NEL
    editor's standard draft~\cite{w3c_nel}.
    Our experiments confirm that current Google Chrome reports the URL parameters in NEL reports.

  \item A dynamic IP address is personal data if there is a mechanism
    that identifies the person behind the IP address~\cite{cjeuC582_14}.
    Up to 60--70\,\% of HTTP user-agent strings
    (UAS) can accurately identify hosts in some datasets, and the number rises
    to 80\,\% when augmented with an IP address~\cite{uas_uniqueness}. Note that
    the UAS research~\cite{uas_uniqueness} is old, and today numbers are likely
    lower as browser vendors limit the information in UASes. Even so,
    later research revealed a long tail of UASes that may reveal
    users~\cite{uas_analysis}.

\item The latest NEL editor's standard draft~\cite{w3c_nel} allows operators to request
  embedding additional headers in NEL reports by signaling optional {\tt
    request\_headers} and {\tt response\_headers} policy members. Personal
    data can be a part of additional HTTP headers reported to the
  server, such as ETag~\cite{WebTrackingPolicyTechnology}, if the original
    server employs such headers for tracking.

\end{enumerate}

Whenever the information collected by NEL consists of personal data, the operator of
the original sever needs to apply additional provisions stemming from GDPR, like
the principle of data minimization, transparency, and accountability. For
example, the minimization principle requires personal data to be \emph{adequate,
relevant and limited to what is necessary in relation to the purposes for which
they are processed}~\cite[Art. 5(1)(c)]{gdpr}. As the original paper did not
find much use in reporting URL paths~\cite[Section 3.4]{paper_nel}, the
controller processing URLs with personal data by NEL would have hard-time
defending the necessity and relevancy of the processing.

Moreover,
the NEL collector operator needs to follow the rules for processors or
joint controllers, including accepting a legal contract.
As the NEL collector can deploy its own NEL policy, operators of the NEL
meta report collectors are also
processors or joint controllers of the original web
server~\cite{edpb_controller} as they process the IP addresses of the
users~\cite{cnil_ga_onestopshop,cjeuC582_14}.
Hence, they need
to conclude a legal contract as well.

\begin{figure}[h!]
  \centering
    \includegraphics[width=0.25\textwidth]{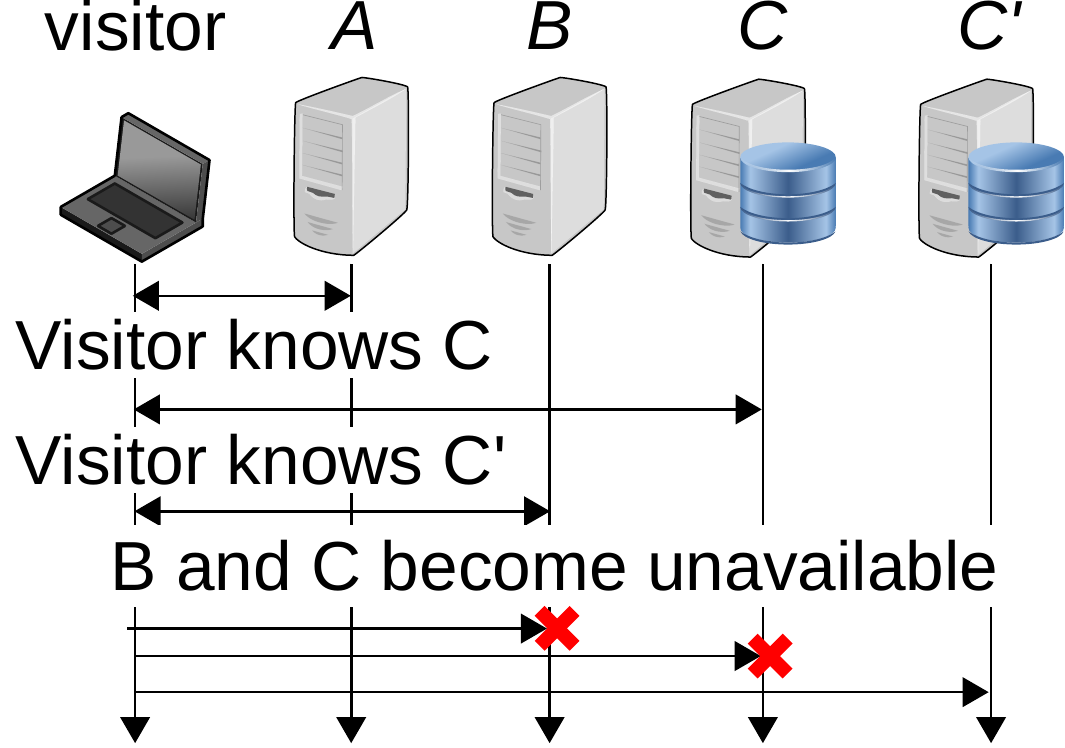}
  \caption{The browser reports an error of $C$.}
 \label{fig:neljoint1}
\end{figure}

The processors or joint controllership brings further obstacles to the
operators of collectors that add NEL headers. Consider that a group
of distinct web servers share a single collector $C$ that adds a NEL
header. Furthermore, suppose that the visitor already reported from web server
$A$ to $C$, so the NEL policy for $C$ to report to $C'$ is already installed in the browser. Later,
the same visitor visits $B$ and installs the NEL policy from $B$ to report to $C$.
Afterwards, both $B$ and $C$ become inaccessible. Consequently, the browser
will report an error to $C'$. Figure~\ref{fig:neljoint1}
shows the communication diagram.

\begin{figure}[h!]
  \centering
    \includegraphics[width=0.4\textwidth]{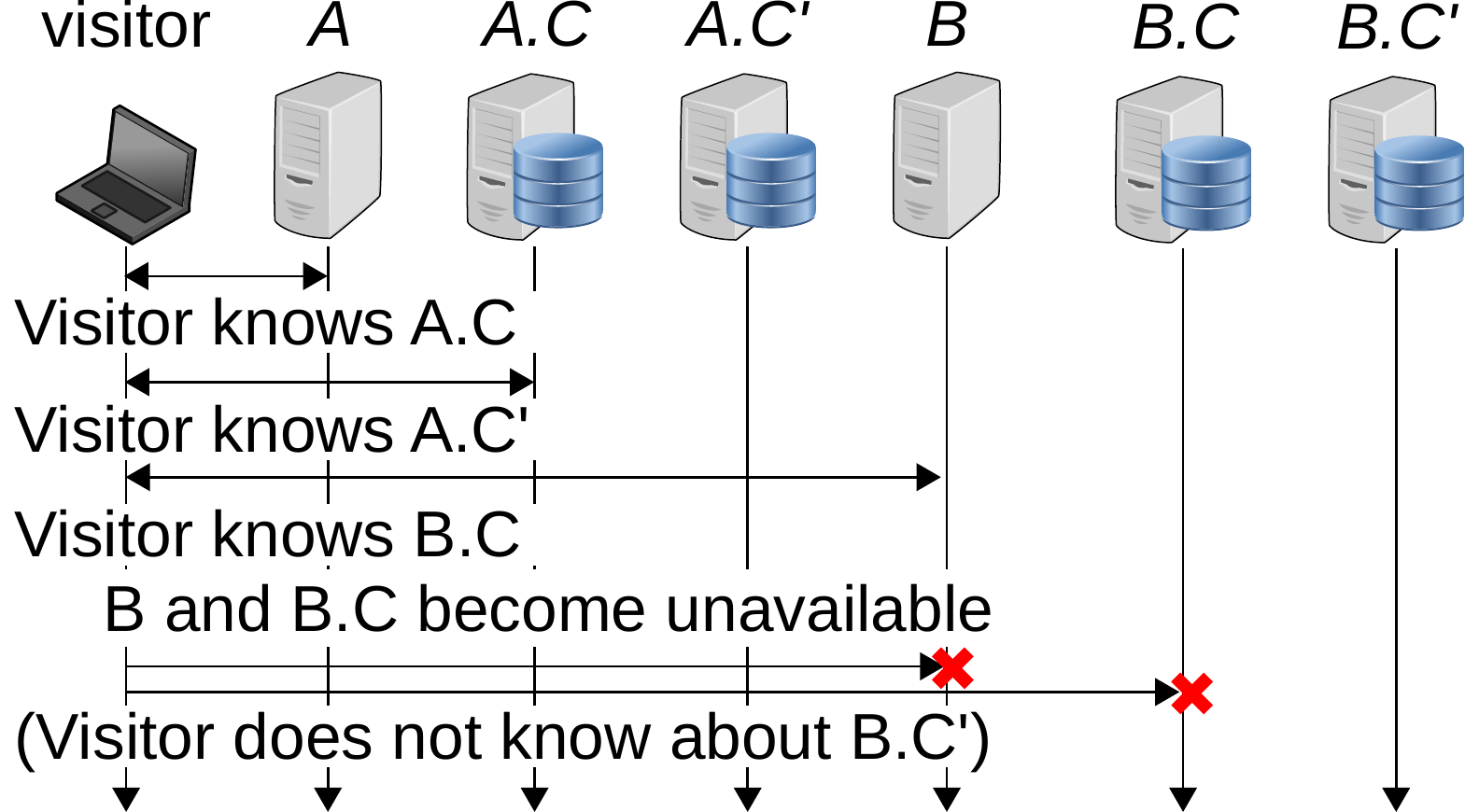}
  \caption{The browser cannot report an error of $B.C(.example)$.}
 \label{fig:neljoint2}
\end{figure}

However, as the browser learned the policy for $C$ through communication
with $A$, both $A$ and $B$ need to share the same joint controllers and
processors. That is indeed legal. However, both $A$ and $B$ must have the
possibility to remove processors or joint controllers from the processing
chain~\cite{edpb_cloud_cef}. Technically, different chains of collectors
can be formed by operating the collectors for
$A$ and $B$ with different domain names;
for example, $A.C.example$ and $B.C.example$.
The downside of such
an approach is that the browser cannot learn the collector of $B.C.example$
through interaction with $A$ and its collectors, see
Figure~\ref{fig:neljoint2}. Consequently, should web server $B$ and its
collector $B.C.example$ not be accessible simultaneously (modified scenario from
above in Figure~\ref{fig:neljoint2}), the browser would not report the problem
with $B.C.example$.

Articles 12-14 of GDPR list requirements on the information that a data
controller should reveal to each data subject before the data processing starts
or in a reasonable time afterwards. Processing
in the scope of ePrivacy Directive yields the same requirements on
transparency (ePrivacy Art. 5(3), GDPR Art. 94(2)).
Hence, each controller employing NEL needs to reveal
transparent information about data collection, for example, in the privacy
policy.

\section{\uppercase{NEL Security, Privacy, And Ethics Analysis}}
\label{sec:disputes}
\noindent The original NEL paper~\cite[Section 3.2]{paper_nel} explains four
security, privacy, and ethical principles. This section disputes the NEL design
and argues that the principles are not fulfilled.
Furthermore, the current NEL editor's standard draft~\cite{w3c_nel}
does not remedy the finding of this section.

\subsection{Can NEL Data Leak to Third Parties?}

Principle 1 establishes that NEL stores existing information in a different
place. NEL policy determines the storage. The 4th principle expects that third
parties must not monitor users through NEL. Hence, third parties should not
be able to install their collectors. Nevertheless, the same paper
\cite[Section~3.3]{paper_nel} warns that the assumption holds only without any
tampering with the certificate validation mechanism. The paper rightly points
out that an adversary having the power to circumvent the validation mechanism
can tamper with the HTTP data by other means, for example, by injecting
additional JavaScript.

However, the original paper, as well as the current NEL
editor's standard draft~\cite{w3c_nel}, does not highlight the persistence of the NEL
policy.
The injection capabilities of the adversary are effective only
during the time that the adversary is in the MitM position. Although the browser can cache the
malicious content and the JavaScript injection attack can be effective some time after the adversary
leaves the MitM position, the cached content is likely to be replaced by updates
on the server or evicted due to the limited size of the cache.

Nevertheless, adversaries with the ability to insert their JavaScript code can
also deploy Service
Workers~\cite{w3c_serviceworkers} that can act as a
proxy between the visited page and its server.
However, the scope of Service Workers is limited
to subpaths of the path where their worker script is located.
Consequently, a Service Worker sees all requests and
replies of the subpath on the web server, can initiate requests, and can influence the content of the
replies.

Similarly, a NEL policy stays on the
device for the whole duration specified by the policy unless replaced by a new
policy sent by the server. The current NEL
editor's standard draft~\cite{w3c_nel} does not limit the maximal duration of the policy, so it
can effectively be installed for eternity (for example, hundreds of years). For
domains that do not signal NEL, the maliciously installed
policy would be removed only when the user clears data stored in the browser.
Unlike Service Workers, NEL policies do not have scope limitations.

Neither the original paper nor the current NEL editor's standard
draft~\cite{w3c_nel} consider adversaries that
appear on a minority of web servers that allow users
to inject their content, like blogs or personal web pages.
For example, such a user can create their content on a specific URL path on the
server.
Some of these servers allow users to modify HTTP headers.
Such \emph{rogue content creators} can deploy NEL (as well as Service Workers) on their part of the web server.

A naïve assumption is that NEL does not increase the powers of
an adversary and only brings an
additional tool. However, we identified three scenarios where NEL creates
additional risks compared with Service Workers:

\begin{itemize}

  \item A NEL policy applies to all pages on the domain that deploys NEL (and
    possibly subdomains). In contrast, the scope of Service Worker is limited.
    Hence, NEL increases the powers of \emph{rogue content creators} that
    control only limited paths on the server as their NEL collector would see
    all browsing activities of the visitors with the web server after their
    visit of the malicious part of the web server. In contrast, Service Worker
    would only see HTTP traffic concerning the subpath of the web server.

  \item Some browser extensions block JavaScript or Service
    Workers~\cite{jshelter,js0}\footnote{For example,  Google Store reports that more
  than 100,000 users installed
    NoScript
    (\url{https://chrome.google.com/webstore/detail/noscript/doojmbjmlfjjnbmnoijecmcbfeoakpjm}).}.
  We tested NoScript 11.4.16 in Chrome, and the extension does not block NEL.
Consequently, NEL allows adversaries like TLS
  MitM proxies and
  rogue content creators to sniff on traffic of users protected from Service
  Workers.

  \item Suppose that an adversary finds a bug that allows to inject
HTTP headers and, thus, NEL
headers. Even if a single page on a domain was affected by such a bug, the adversary would be
able to monitor the traffic of the visitors of the vulnerable page on the whole
domain and its subdomains. Note that the hypothetical bug does not require that the
attacker has the ability to run JavaScript code in the visiting browsers that is
needed to install Service Workers.

\end{itemize}

Neither the original paper nor  the current NEL
editor's standard draft~\cite{w3c_nel} discuss such or similar
scenarios.

Web server operators that deploy NEL do not suffer from the risk as the
legitimate policy replaces the malicious policy during the next visit.
We recommend that other web server operators add policies with
{\tt max\_age=0}. The
duration of 0 forces the browser to remove any previously stored NEL policy for
the domain, including the policies adversaries may have inserted in
the past. \cite{nel_http_archive} reports less than ten domains removing NEL
policies in each analyzed year, so we conclude that we are the first to
promote the advice publicly.

We believe that signalling NEL policy removal does not trigger the legal issues
identified in Section~\ref{sec:law} with NEL in
EEA, as the recommended policy would not remove any NEL policy under normal
circumstances. Hence, ePrivacy Article 5(3) would not be triggered. A malicious
policy would not be requested by the user, so it is not necessary to provide the
service. As the web server operator has to process personal data securely (GDPR,
Art. 32), ensuring that browsing history does not leak to an adversary is in the spirit
of the law.

\subsection{Does NEL Provide Data Only When Users Voluntarily Access The Service?}

A user or a network operator can deploy DNS firewall, for example, by
changing the \emph{host} file or the local DNS resolver~\cite{dns_firewall_muni}.
Consequently, the DNS firewall returns invalid IP addresses for domain
queries of the blocked domains. So, for example, if a web page $A$ includes
content from a blocked domain $D$, the browser cannot access the server of $D$ as the DNS returns an
invalid IP address.

However, NEL policies installed before the deployment of such a DNS firewall stay in
place. Hence, the browser would report to the collector of the blocked
domain (if the collector is not also blocked) that the IP address of the server changed,
the IP address to which the DNS firewall remaps the domain, and the IP address
of the computer running behind the DNS firewall.

We argue that the user or the operator that deployed the DNS firewall took active
measures against accessing the blocked domains. Despite this, NEL reveals the
DNS firewall to the blocked party.

Another scenario where NEL can be exploited to report to services that users do
not access voluntarily is online advertisement. Some data protection authorities~\cite{ico_adtech} report that
people are often not aware of the queries of their browsers interacting with
online advertisement business. Suppose that one of these companies deploys NEL. We argue that if the data protection
authorities claim that the online advertisement processing is \emph{"disproportionate, intrusive, and
unfair"}~\cite{ico_adtech}, users do not voluntarily access such services.
Hence, in this scenario, NEL would track HTTP services that the
users are unaware of and thus do not access voluntarily.

\subsection{Can End Users Opt Out of NEL?}

Both the original NEL paper~\cite{paper_nel} and the current NEL
editor's standard draft~\cite{w3c_nel} expect that the
users can opt out of NEL. However, we
cannot find any option to disable NEL in Google Chrome 110 either
in settings or in {\tt chrome://flags}, even though we went through the settings option-by-option.
Moreover, we tried to seek help from Google Search. However, the engine did not return any
result for queries \emph{"chrome disable network error logging"} and
\emph{"chrome disable nel"} neither on the whole web nor on
\url{stackoverflow.com}; for each query, we went through the first 50 results
offered by the search engine. Even more, we asked a long-term Chrome user who is
a Ph.D. student of IT to help us. He was not able to deactivate NEL during 15
minutes of active trying.

We argue that if Chrome offers a setting to disable NEL, it is hidden and cannot
be easily located by regular users.

\section{\uppercase{Discussion}}
\label{sec:discussion}
\noindent This section discusses the results %and limitations
of this paper.

\subsection{A Ticking Data Protection Bomb}
%\label{sec:bomb}

\cite{nel_http_archive} report that more than 11.7\,\% of domains
deploy NEL. Yet, Section~\ref{sec:law} argues that users in EEA need
to consent to NEL. Consequently, the operators of the web servers deploying NEL
may be in conflict with the law. However, the number of
11.7\,\% is an upper bound. Not all of the domains are in the scope of the
ePrivacy Directive. Additionally, previous work~\cite{ePr_4years} shows
a heterogeneous implementation of the ePrivacy Directive by the EU (European
Union) member states.
Hence, some operators in the EU might not violate the law due to the incorrect
implementation of the ePrivacy Directive in their state.

The plan is to replace ePrivacy Directive with a Regulation\footnote{See the
procedure and related documents at
\url{https://eur-lex.europa.eu/legal-content/EN/HIS/?uri=CELEX:52017PC0010}}.
Initially, the plan was to replace the Directive on the same day as GDPR
entered into force. Yet, as of the writing of this paper in February 2023, the
final text of the new Regulation is not known. Neither the text proposed by The
European Commission nor the proposed amendments by The European Parliament (EP) change
the requirement for consent to store NEL policy in a browser. However,
The Council of the European Union (Council) version
provides amendment Article 8(1)(da) that would allow using storage
capabilities of terminal equipment to detect technical faults for the duration
necessary.

The authors of this paper interpret the Council Article 8(1)(da) as allowing
NEL policies with 0 success fraction, provided that the operator erases the logged data
when each failure is detected and mitigated.

As of the time of the writing of this text, it is not clear if an exception like
the Council Article 8(1)(da) propagates to the final ePrivacy Regulation. Once
the Council and the EP agree on the final text of the Regulation,
there should be a 2-year transition period before the new Regulation
takes effect.

\subsection{Cloudflare}

Cloudflare is the largest NEL collector provider~\cite{nel_http_archive}.
Let us focus on their data protection information.

The public information on NEL support is confusing. The \emph{Get started guide}
looks like the web operators need to explicitly enable
NEL\footnote{\url{https://developers.cloudflare.com/network-error-logging/get-started/}}.
However, on another page, Cloudflare informs its customers that they \emph{"can opt out
of having their end users consume the NEL headers by emailing Cloudflare
support"}\footnote{\url{https://support.cloudflare.com/hc/en-us/articles/360050691831-Understanding-Network-Error-Logging}\label{ft:understand_nel}}.
Some customers report that they cannot see NEL analytics on their
dashboard\footnote{\url{https://community.cloudflare.com/t/access-to-network-error-logging/472139}}.
The customers speculate that Cloudflare uses NEL data internally.

Cloudflare stores IP
addresses only in volatile
memory to estimate the physical and network location of the user and determine
if the user changed the IP address recently\footnote{See footnote \ref{ft:understand_nel}.}.
It seems that Cloudflare interprets the law that the volatility of the storage
means that it does
not process personal data. However, the French data protection
authority~\cite{cnil_ga_onestopshop} recently treated IP addresses in Google
Analytics as personal data. Even though Google Analytics trims the addresses
shortly after the arrival, there is a period when full addresses are available, and
consequently, personal data are processed.

If Cloudflare enabled NEL without the
awareness of their customers, Cloudflare would be guilty of breaching the
ePrivacy Directive and likely breaching GDPR Art. 28(3)(a) as it processes
personal data without such instructions of the controller during processing that is
in the scope of the supposedly breached EEA law.

An analysis of other NEL players' data protection issues is out
of the scope of this paper as
each deployment needs to be analyzed individually. Furthermore, we are not Cloudflare
customers, so the data protection analysis did not consider information
agreed upon between Cloudflare and its customers, like the contracts.

\section{\uppercase{Conclusion}}
\label{sec:conclusion}
\noindent Error monitoring is a crucial activity of web server operators.
Even though Chromium-based browsers have supported NEL for several years,
the ongoing NEL
standardization
received little academic scrutiny. To our best knowledge, neither the European
Data Protection Board nor its predecessor WP29
issued any opinion or guidelines about NEL.

This paper identified several NEL deficiencies from the
European data protection view. First of all, the authors of this paper interpret
the law as requiring consent before an operator can install a NEL policy. However,
NEL is supposed to be opt-out~\cite{paper_nel,w3c_nel}, so there is no built-in
way to seek consent (opt-in). Moreover, web operators need to be cautious about
personal data reported by NEL as they trigger additional requirements on data
minimization, transparency, and accountability.

Although the security, privacy, and ethics analysis of the original
paper~\cite{paper_nel} supposes that NEL data cannot leak to third parties, NEL
provides data only when users voluntarily access the service, and end users can
opt out of NEL, our paper provides counter-arguments explaining that the
expectations are not valid in all cases. Most strikingly, we failed to
opt out of NEL in Chrome.

\section*{\uppercase{Acknowledgments}}
This work was supported by the Brno University of
Technology grant Smart information technology for a resilient society
(FIT-S-23-8209).

\bibliographystyle{apalike}
{\small
\bibliography{biblio}}

\begin{thebibliography}{}

\bibitem[Acar et~al., 2020]{session-replay}
Acar, G., Englehardt, S., and Narayanan, A. (2020).
\newblock No boundaries: data exfiltration by third parties embedded on web
  pages.
\newblock {\em Proceedings on Privacy Enhancing Technologies}, 2020:220--238.

\bibitem[Burnett et~al., 2020]{paper_nel}
Burnett, S., Chen, L., Creager, D.~A., Efimov, M., Grigorik, I., Jones, B.,
  Madhyastha, H.~V., Papageorge, P., Rogan, B., Stahl, C., and Tuttle, J.
  (2020).
\newblock Network error logging: Client-side measurement of end-to-end web
  service reliability.
\newblock In {\em 17th {USENIX} Symposium on Networked Systems Design and
  Implementation, {NSDI} 2020}, pages 985--998. {USENIX} Association.

\bibitem[{Commission Nationale de l’Informatique et des Libertés},
  2022]{cnil_ga_onestopshop}
{Commission Nationale de l’Informatique et des Libertés} (2022).
\newblock {EDPBI:FR:OSS:D:2022:330}.
\newblock
  \url{https://edpb.europa.eu/system/files/2022-08/fr_2022-02_decisionpublic_redacted_0.pdf}.

\bibitem[Conti et~al., 2016]{mitmsurvey}
Conti, M., Dragoni, N., and Lesyk, V. (2016).
\newblock A survey of man in the middle attacks.
\newblock {\em IEEE Communications Surveys \& Tutorials}, 18(3):2027--2051.

\bibitem[{Court of Justice of the European Union}, 2016]{cjeuC582_14}
{Court of Justice of the European Union} (2016).
\newblock {Case C‑582/14: Patrick Breyer v. Bundesrepublik Deutschland}.
\newblock {ECLI:EU:C:2016:779}.

\bibitem[{Court of Justice of the European Union}, 2021]{cjeuC645_19}
{Court of Justice of the European Union} (2021).
\newblock {Case C‑645/19: Facebook Ireland Ltd, Facebook Inc., Facebook
  Belgium BVBA v. Gegevensbeschermingsautoriteit}.
\newblock {ECLI:EU:C:2021:483}.

\bibitem[{Europan Data Protection Board}, 2021]{edpb_controller}
{Europan Data Protection Board} (2021).
\newblock Guidelines 07/2020 on the concepts of controller and processor in the
  {GDPR}.
\newblock
  \url{https://edpb.europa.eu/system/files/2021-07/eppb_guidelines_202007_controllerprocessor_final_en.pdf},
  Version 2.1.

\bibitem[{Europan Data Protection Board}, 2023]{edpb_cloud_cef}
{Europan Data Protection Board} (2023).
\newblock { 2022 Coordinated Enforcement Action - use of cloud-based services
  by the public sector }.
\newblock
  \url{https://edpb.europa.eu/system/files/2023-01/edpb_20230118_cef_cloud-basedservices_publicsector_en.pdf}.

\bibitem[Franke, 2012]{franke2012}
Franke, U. (2012).
\newblock Optimal {IT} service availability: Shorter outages, or fewer?
\newblock {\em IEEE Transactions on Network and Service Management},
  9(1):22--33.

\bibitem[{Information Commissioner’s Office}, 2019]{ico_adtech}
{Information Commissioner’s Office} (2019).
\newblock Update report into adtech and real time bidding.
\newblock
  \url{https://ico.org.uk/media/about-the-ico/documents/2615156/adtech-real-time-bidding-report-201906.pdf}.

\bibitem[Jeřábek and Polčák, 2023]{nel_http_archive}
Jeřábek, K. and Polčák, L. (2023).
\newblock Network error logging: {HTTP} archive analysis.
\newblock Preprint at \url{https://arxiv.org/abs/2305.01249}.

\bibitem[Kline et~al., 2017]{uas_analysis}
Kline, J., Barford, P., Cahn, A., and Sommers, J. (2017).
\newblock On the structure and characteristics of user agent string.
\newblock In {\em Proceedings of the 2017 Internet Measurement Conference},
  page 184–190, New York, NY, USA. Association for Computing Machinery.

\bibitem[Krishnamurthy and Wills, 2011]{Krishnamurthy2011}
Krishnamurthy, B. and Wills, C. (2011).
\newblock Privacy leakage vs. protection measures: the growing disconnect.
\newblock In {\em Proceedings of the Web 2.0 Security and Privacy Workshop}.

\bibitem[Li et~al., 2018]{hsts_survey}
Li, X., Wu, C., Ji, S., Gu, Q., and Beyah, R. (2018).
\newblock {HSTS} measurement and an enhanced stripping attack against {HTTPS}.
\newblock In {\em Security and Privacy in Communication Networks}, pages
  489--509, Cham. Springer International Publishing.

\bibitem[{Mayer} and {Mitchell}, 2012]{WebTrackingPolicyTechnology}
{Mayer}, J.~R. and {Mitchell}, J.~C. (2012).
\newblock Third-party web tracking: Policy and technology.
\newblock In {\em 2012 IEEE Symposium on Security and Privacy}, pages 413--427.

\bibitem[Michael~Schwarz and Gruss, 2018]{js0}
Michael~Schwarz, M.~L. and Gruss, D. (2018).
\newblock Javascript zero: Real javascript and zero side-channel attacks.
\newblock In {\em Network and Distributed Systems Security Symposium 2018}.

\bibitem[Polčák et~al., 2023]{jshelter}
Polčák, L., Saloň, M., Maone, G., Hranický, R., and McMahon, M. (2023).
\newblock {JShelter}: Give me my browser back.
\newblock In {\em Proceedings of the 20th International Conference on Security
  and Cryptography}. SciTePress - Science and Technology Publications.

\bibitem[Raman et~al., 2020]{kazakhstan_ca}
Raman, R.~S., Evdokimov, L., Wurstrow, E., Halderman, J.~A., and Ensafi, R.
  (2020).
\newblock Investigating large scale https interception in {Kazakhstan}.
\newblock In {\em Proceedings of the ACM Internet Measurement Conference}, IMC
  '20, pages 125--132, New York, NY, USA. Association for Computing Machinery.

\bibitem[Stricot-Tarboton et~al., 2016]{https_mitm_vulnerabilities}
Stricot-Tarboton, S., Chaisiri, S., and Ko, R.~K. (2016).
\newblock Taxonomy of man-in-the-middle attacks on {HTTPS}.
\newblock In {\em 2016 IEEE Trustcom/BigDataSE/ISPA}, pages 527--534.

\bibitem[Sunshine et~al., 2009]{ssl_effectiveness}
Sunshine, J., Egelman, S., Almuhimedi, H., Atri, N., and Cranor, L.~F. (2009).
\newblock Crying wolf: An empirical study of {SSL} warning effectiveness.
\newblock In {\em Proceedings of the 18th Conference on {USENIX} Security
  Symposium}, SSYM'09, pages 399--416. USENIX Association.

\bibitem[{The European Parliament and The Council of the European Union},
  2002]{ePrivacy}
{The European Parliament and The Council of the European Union} (2002).
\newblock {Directive 2002/58/EC. Directive on privacy and electronic
  communications}.
\newblock {Official Journal L 201}.

\bibitem[{The European Parliament and The Council of the European Union},
  2009]{ePrivacyAmmendment}
{The European Parliament and The Council of the European Union} (2009).
\newblock {Directive 2009/136/EC}.
\newblock {Official Journal L 337}.

\bibitem[{The European Parliament and The Council of the European Union},
  2016]{gdpr}
{The European Parliament and The Council of the European Union} (2016).
\newblock {Regulation 2016/679/EU. General Data Protection Regulation (GDPR)}.
\newblock {Official Journal of the European Union L 119, 4.5.2016}.

\bibitem[{The Office for Personal Data Protection}, 2021]{UOOUReservatics}
{The Office for Personal Data Protection} (2021).
\newblock Statement on the reservation system for covid-19 vaccination.
\newblock
  \url{https://www.uoou.cz/en/vismo/dokumenty2.asp?id_org=200156&id=2011}.

\bibitem[Trevisan et~al., 2019]{ePr_4years}
Trevisan, M., Travers, S., Bassi, E., and Mellia, M. (2019).
\newblock 4 years of {EU} cookie law: Results and lessons learned.
\newblock {\em Proceedings on Privacy Enhancing Technologies}, 2019:126--145.

\bibitem[van Hoboken and Zuiderveen~Borgesius, 2015]{vanhoboken2015}
van Hoboken, J. and Zuiderveen~Borgesius, F. (2015).
\newblock Scoping electronic communication privacy rules: Data, services and
  values.
\newblock {\em Journal of Intellectual Property, Information Technology and
  Electronic Commerce Law}, 6(3):198--210.

\bibitem[{W3C}, 2021]{w3c_nel}
{W3C} (2021).
\newblock Network error logging.
\newblock The World Wide Web Consortium (W3C) internal document,
  \url{https://w3c.github.io/network-error-logging/}, Editor's Draft of 30th
  July 2021.

\bibitem[{W3C}, 2022a]{w3c_serviceworkers}
{W3C} (2022a).
\newblock Push {API}.
\newblock The World Wide Web Consortium (W3C), Web Applications Working Group,
  \url{https://www.w3.org/TR/push-api/}, Working Draft of 30 June 2022.

\bibitem[{W3C}, 2022b]{w3c_push}
{W3C} (2022b).
\newblock Service workers nightly.
\newblock The World Wide Web Consortium (W3C), Service Workers Working Group,
  \url{https://w3c.github.io/ServiceWorker/}, Editor's Draft of 28 November
  2022.

\bibitem[{WP29}, 2012]{wp29_cookies_consent}
{WP29} (2012).
\newblock {Opinion 04/2012 on Cookie Consent Exemption}.
\newblock Article 29 Data Protection Working Party, WP 194,
  \url{https://ec.europa.eu/justice/article-29/documentation/opinion-recommendation/files/2012/wp194_en.pdf}.

\bibitem[Yen et~al., 2012]{uas_uniqueness}
Yen, T.-F., Xie, Y., Yu, F., Yu, R.~P., and Abadi, M. (2012).
\newblock Host fingerprinting and tracking on the web: Privacy and security
  implications.
\newblock In {\em Network and Distributed System Security (NDSS) Symposium}.

\bibitem[Špaček et~al., 2019]{dns_firewall_muni}
Špaček, S., Laštovička, M., Horák, M., and Plesník, T. (2019).
\newblock Current issues of malicious domains blocking.
\newblock In {\em 2019 IFIP/IEEE Symposium on Integrated Network and Service
  Management (IM)}, pages 551--556.

\end{thebibliography}

\end{document}